\documentclass[12pt]{article}
\usepackage{fancyhdr}
\usepackage{authblk}
\usepackage[square,numbers]{natbib}
\usepackage[letterpaper,margin=1in,bmargin=1in]{geometry}
\usepackage{amsthm}
\usepackage{wrapfig}
\usepackage{array}
\usepackage{color}
\usepackage{url}
\usepackage[dvipsnames]{xcolor}
\usepackage[cmex10]{amsmath} 
\usepackage{amssymb}
\usepackage{bbm} 
\usepackage{graphicx} 
\usepackage{enumitem}
\usepackage{amsmath}
\usepackage{amsfonts}
\usepackage{cleveref}
\usepackage{xspace}

\newtheorem{example}{Example}

\newcommand{\prop}{{\textsf{prop}}\xspace}
\newcommand{\props}{{\textsf{props}}\xspace}
\newcommand{\Props}{{\textsf{Props}}\xspace}

\title{\Props for Machine-Learning Security}
\author[1]{Ari Juels} \author[2]{Farinaz Koushanfar}
\affil[1]{Cornell Tech}
\affil[2]{University of California San Diego}
\date{\today}

\begin{document}

\maketitle

\begin{abstract}
We propose \textit{protected pipelines} or \textit{\props} for short, a new approach for authenticated, privacy-preserving access to deep-web data for machine learning (ML). By permitting secure use of vast sources of deep-web data, \props address the systemic bottleneck of limited high-quality training data in ML development. \Props also enable privacy-preserving and trustworthy forms of inference, allowing for safe use of sensitive data in ML applications. Finally, \props offer a new approach to constraining adversarial inputs. \Props are practically realizable today by leveraging privacy-preserving oracle systems initially developed for blockchain applications.
\end{abstract}

\section{Introduction}

There’s only one World Wide Web (WWW). This simple fact represents a major barrier for advances in machine learning, as practitioners are reaching fundamental limits in available data sources for model training~\cite{longpre2024consent,villalobosposition}. They’re relying increasingly instead on synthetic data, a partial remedy that carries the risk of training models that are self-poisoned or misrepresent the real world~\cite{nikolenko2021synthetic}. Practitioners are also making efforts to tap private sources of data. Obtaining access to sensitive sources of data, though, often requires legally complex and labor-intensive negotiations over conditions of use and subjects large populations of users to the risk of privacy breaches.

Fortunately, there is more than one WWW. There is the \textit{surface web}, with its publicly accessible, indexed data. Then there is the \textit{deep web}. The deep web---meaning data sources walled-off from scraping, and ranging from legitimate consumer and enterprise environments to (dark-web) sites of illicit activity---is estimated to be two orders of magnitude larger than the surface public web~\cite{bergman2001white,rai2020bibliometric}. 

Deep-web data can include personal data such as e-mail, health or fitness data from personal devices or medical providers, e-mail, digital calendars, photographs, and financial statements. It can also include documents maintained by organizations, including billing and accounting data, customer orders, transaction records and so forth.
As we explain, little of such data can be shared securely with today's web infrastructure, greatly limiting its availability for ML applications. 

In this short work, we propose a new idea that we call \textit{protected pipelines} for machine learning security or \textit{\props} for short. \Props enable secure access to deep web data. The security assurances that \props provide---which include both integrity and privacy---also give rise to both new security models for ML applications and new models of data sharing for end users. Critically, \props \textit{do not require any modification to existing web infrastructure}.

We give an overview of \props in~\Cref{sec:what}, along with examples to illustrate their applications. In~\Cref{sec:built}, we explain how they can be built using privacy-preserving oracles designed for blockchain systems. We summarize our ideas in~\Cref{sec:conclusion}. Our examples focus on private user data, as   secure sharing of such data is especially challenging given today's web infrastructure. 

\section{What Are \Props?}
\label{sec:what}

\Props are a data pipeline extending from deep-web data sources to points of use in the ML ecosystem. They enforce two key security properties. 

The first property of \props is \textit{privacy}, specifically the ability of a user to retain control over disclosure of data throughout the pipeline. \Props may be viewed as enforcing a common notion of privacy known as ``contextual integrity,'' meaning that data flows appropriately according to its intended use~\cite{barth2006privacy,nissenbaum2010privacy}. \Props in particular ensure that data remain confidential in the greatest degree that is consistent with their use in target applications.

The second property of \props is \textit{integrity}, meaning specifically that \props prove to consumers of deep-web data and users of downstream models relying on these data that the data are authentic, i.e., come from trustworthy deep-web sources.

\Props can support both model \textit{training} and \textit{inference}.

\subsection{\Props for model training}
Here is an illustration of the use of \props for model training.

\begin{example}[Training: Health data]
\label{ex:training}
MediModels Inc. is training a health-diagnostics ML model. Alice wishes to furnish her electronic health record (EHR) as training data for the model.
    
Alice could just download her EHR—let’s denote it by $X$---from her medical provider BigHospital (e.g., as a PDF) and send it to MediModels. But then MediModels would have no way to ensure that $X$ is real, i.e., not modified or fabricated by Alice. Fake data from malicious users (or competitors) could irremediably corrupt MediModel’s model. 

Alice can instead use an app, provided by MediModels, that realizes a \prop for ML. This app enables Alice to log into the web portal of BigHospital, obtain her EHR, and then relay her EHR to MediModels. 

MediModels obtains high assurance that $X$ is authentic: it is the result of Alice sourcing her EHR from BigHospital. Alice consents to and controls release of her information.
Critically here, in our proposed approach BigHospital \textbf{need not modify its web servers or even know of the use or existence of MediModels’ app}.
\end{example}

\Props are compatible with privacy-preserving ML training systems, such as federated learning~\cite{kairouz2021advances} or use of trusted execution environments~\cite{nertney_confidential_computing}. In~\Cref{ex:training}, if MediModels is using such a system, Alice’s data would never be disclosed explicitly to MediModels or anyone else. It would only serve for model training.\footnote{There is significant literature on the extraction of training data from models (see, e.g.,~\cite{nasr2023scalable,song2017machine}), so caution is still needed even when the process of model training itself operates on confidential inputs. 
}

\begin{figure}[!ht]
    \centering
    \includegraphics[width=0.7\linewidth]{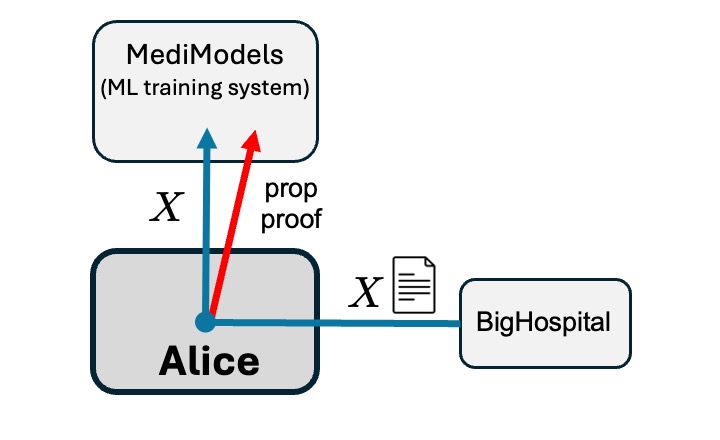}
    \caption{Illustration of Example 1. Alice obtains her EHR $X$ from BigHospital and relays it to MediModels. A \prop proof shows that $X$ is authentic, the result of Alice querying BigHospital's web portal for her EHR.} 
    \label{fig:enter-label}
\end{figure}

\subsection{\Props for inference}

\Props can also support pipelines for ML inference. In this case, a \prop proves that an inference results from applying a particular model to authenticated, sensitive source data---without directly revealing the data. In other words, it provides precise \textit{provenance} for an inference result. \Cref{ex:inference} illustrates this idea. We use the informal notation $M(X)$ to denote the application of a model $M$ to source data $X$.

\begin{figure}[!ht]
    \centering
    \includegraphics[width=0.7\linewidth]{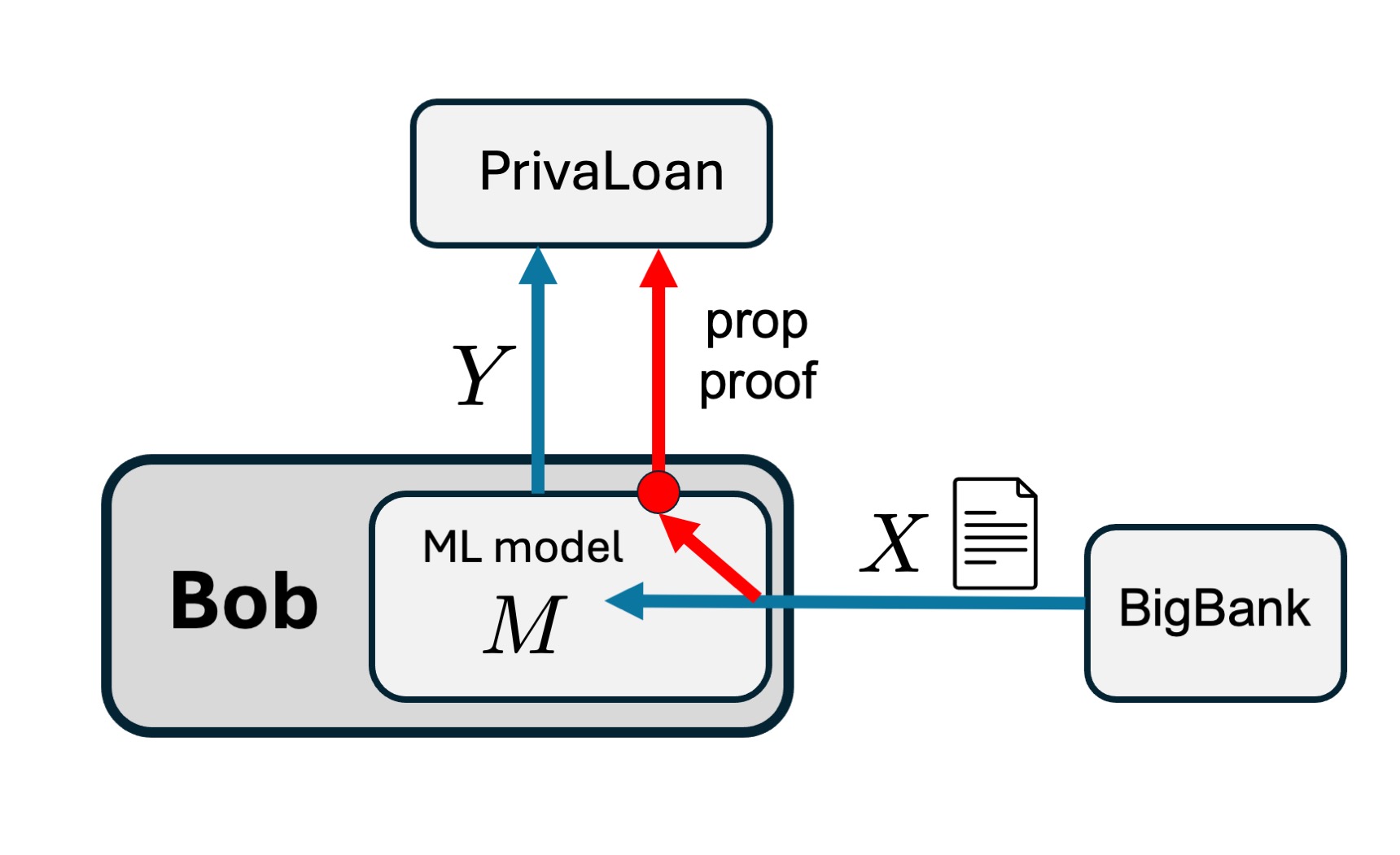}
    \caption{Illustration of Example 2. Bob obtains a financial document $X$ from BigBank. He runs model $M$ on it and sends the output $Y$ to PrivaLoan. A \prop proof shows that $Y = M(X)$ for an authentic document $X$.}
    \label{fig:bob-privaloan}
\end{figure}

\begin{example}[Inference: Privacy-preserving loan decision]
    \label{ex:inference}

Bob applies for a loan from a new financial services company called PrivaLoan. PrivaLoan has an innovative approach to lending. Bob obtains a set $X$ of trustworthy financial documents (e.g., transaction statements) from any of a range of pre-approved sources (major banks and brokerage firms) on the web, e.g., BigBank. He then uses a \prop-enabled PrivaLoan app to: (1) Execute a PrivaLoan loan-decision model $M$ on $X$ on his own mobile phone, resulting in loan decision $Y$ and then (2) Generate a proof showing that $Y = M(X)$ for $X$ a set of validly sourced documents. PrivaLoan then acts on decision $Y$. \Cref{fig:bob-privaloan} illustrates this \prop inference scenario.
\end{example}

This example illustrates how a consumer can use sensitive data privately to support organizations’ decision-making. Such privacy-preserving inference is not just of benefit to users, but also enables organizations to avoid the risks of handling sensitive consumer data and could position stakeholders to align with or even set the standards for privacy-first ML applications in order to meet regulations such as HIPAA and GDPR.

It is also possible to apply \props here to surface internet data, i.e., public data. In this case, data privacy isn't at issue, but integrity remains important, i.e., assurance that $Y = M(X)$ for public data $X$ (or a combination of public and private data) remains valuable as evidence of the trustworthiness of $Y$.

\paragraph{Note (Remote execution):} In~\Cref{ex:inference}, Bob executes $M$ locally. As $M$ is executing on user devices, it is in effect publicly disclosed. An alternative is possible in which $M$ is instead executed in a privacy-preserving cloud environment. In fact, \props compose naturally with a range of privacy-preserving inference systems, such as Apple Intelligence Private Compute Cloud (PCC)~\cite{apple_private_cloud_compute_2023}.

In the case that $M$ isn't operated by the consumer of the inference---e.g., PrivaLoan in~\Cref{ex:inference}---a \prop proof would include two components: (1) A proof of authenticity of $X$ coupled with (2) Proof that $Y = M(X)$.\footnote{While Apple Intelligence PCC doesn't currently generate proofs of the form $Y = M(X)$, it does run on trusted execution environments with attestation capabilities. It could in principle therefore generate such proofs for consumption of inferences by parties other than Apple itself.}

\subsection{Constraining adversarial inputs}

\Cref{ex:inference} illustrates how \props additionally offers a strategy for combating adversarial examples. Adversarial examples are maliciously generated inputs designed to cause models to produce erroneous outputs~\cite{goodfellow2014explaining,ilyas2019adversarial,papernot2016limitations,yuan2019adversarial}. PrivaLoan can trust  $Y$ a because of its trustworthy provenance: \Props authenticate the entire pipeline, from data source to output. As a result, an adversary has a limited ability to manipulate inputs. 

Similarly, by constraining adversarial inputs, \props can serve as a potential countermeasure to other forms of attack, such as model extraction~\cite{tramer2016stealing} and recovery of sensitive training data~\cite{nasr2023scalable,song2017machine}.

\subsection{Data control, monetization, and decentralization}

A user can choose to \textit{pre-process} the data $X$ obtained from a deep-web source for input to a \prop. That is, she can choose to transmit some $X' = f(X)$, where a \textit{filter} $f$ excises or compresses data in $X$. The \prop will then prove---using functionality already available in privacy-preserving oracle systems---that $X'$ is the result of applying $f$ to an authentically sourced $X$. The filter $f$ can redact data or compute over data in order, e.g., to compress or add noise it~\cite{el2022differential} to hedge against privacy failures should data leakage occur downstream.

In~\Cref{ex:training}, for instance, rather than transmitting $X$ to MediModels, Alice might wish to transmit a redacted EHR $X'$ from which she has excised her name and address. (Perhaps she's concerned that these might leak from the trained model.) The \prop specifies the filter $f$ to MediModels. Thus MediModels learns that contact information is omitted from $X'$. MediModels can, of course, choose to accept or reject an input $X'$ based on the filter $f$ that generated it and might, for example, whitelist a set of pre-approved filters. 

In short, users authorize the release of their data in \props and can control this release in a granular way. \Props could support a financial model in which an organization training an ML model compensates users for the data they furnish and filtering choices they apply. 

\Props could also support new, decentralized financial models in which users who provide training data receive a financial stake in a resulting ML model. It is in principle possible, for instance, to train and execute an ML model on a TEE-enabled blockchain such as Oasis Sapphire that ensures data privacy and can automatically bill for queries and distribute cryptocurrency tokens to community members who have earned a stake in the model~\cite{oasis_decentralized_ai}.

\paragraph{Note (Ownership rights):} We don't address the issue of data ownership here. In some cases, the right to make use of personal data as desired is clearly attributable to an individual user and supported by regulations such as Article 20 of GDPR~\cite{GDPR_Article20}, the right to data portability and, in the case of EHRs, the 21st Century Cures Act~\cite{CuresAct_2016}. In other cases, as with photographs of individuals captured in private settings, legal restrictions, e.g.,~\cite{CCPA_2018}, or service agreements may limit a user's sharing rights. It is the responsibility of an application developer to enforce appropriate data sharing policies. The authentication of data sources offered by \props can be instrumental in this goal.

\section{How Can \Props Be Built?}
\label{sec:built}

Why do \props not yet exist? How can we build them? Two critical building blocks are needed: \textit{secure data sourcing} and \textit{pinned models}. Both are practically realizable today using existing tools and techniques.

\subsection{Secure data sourcing} 

Secure data sourcing ensures that the data entering a \prop comes from a trustworthy source, such as a specific web service, in the expected context and with strong privacy protections. In \Cref{ex:training} this means ensuring that $X$ represents BigHospital serving Alice’s EHR. A practical enhancement allows Alice to apply redactions or other preprocessing before transmitting $X$. Alice’s data can then be input to the model-training environment—in encrypted form if the environment is privacy-preserving.
Today, however, secure channels to web servers (TLS / HTTPS) do not digitally sign data~\cite{rfc8446}. That means that while users can access their own web data securely, there’s often no way to prove to someone else where the data came from. There are two ways to remedy this limitation of existing infrastructure.

\paragraph{Approach 1: Infrastructure Modification.}

The first way to address the problem is to change the infrastructure, i.e., modify existing web services so that they sign data. JSON Web Tokens (JWTs) are an emerging standard for this purpose~\cite{jwt_io}. JWTs are gaining traction for certain forms of data, such as user credentials in OAuth 2.0 and OpenID Connect (OIDC). But most deep-web data isn’t served today in the form of JWTs.

\paragraph{Approach 2: Privacy-Preserving Oracles.}

A second, infrastructure-independent approach involves privacy-preserving oracles~\cite{breidenbach2021chainlink}. These are tools developed for blockchain systems that allow secure data sourcing without modifying existing infrastructure. 
Privacy-preserving oracles come in two flavors. They can use trusted execution environments (TEEs) such as Intel SGX / TDX~\cite{cheng2024intel,costan2016intel,mckeen2016intel,mckeen2013innovative} a technology that is increasingly supported in CPUs and even GPUs~\cite{nertney_confidential_computing}. Town Crier~\cite{TownCrier:2018,zhang2016town} was the first such oracle system. TEEs are flexible and powerful, but have long-recognized security limitations, such as repeatedly demonstrated vulnerability to side-channel attacks (e.g., in speculative execution)~\cite{borrello2022aepic,chen2019sgxpectre,kocher2020spectre,lipp2018meltdown,van2018foreshadow,van2020sgaxe,van2024sok}. An alternative is to use a cryptographic alternative, often today called zkTLS~\cite{wetzel2024tls} and first realized in the DECO system~\cite{zhang2020deco}. Both approaches enable a user to furnish deep-web data privately and with integrity to third parties in a \prop as in our two examples. They work with any TLS-enabled web service. They also enable privacy protection as illustrated above: Data $X$ can be pre-processed by a user and sent into a \prop in encrypted form.

\subsection{Pinned models}

Secure data sourcing is generally sufficient for privacy-preserving model training, as in Example 1. But inference is another story. That requires the second building block for \props, pinned models. 

In~\Cref{ex:inference}, it isn’t enough for PrivaLoan, the consumer of the model’s output $Y$, to know that $X$ is trustworthy. PrivaLoan also needs to know what model $M$ was used for inference and the full execution environment $E$ for $M$ (hyperparameters, preprocessing, postprocessing, random seeds, etc.). If $M$ was PrivaLoan's own model, PrivaLoan will naturally want assurance that the inference $Y$ was the output of $M$ on $X$. 

It may suffice for PrivaLoan’s purposes only to have model consistency, i.e., to know that $X$ was input to a particular ML service, such as ChatGPT, without knowing exactly what $E$ and $M$ were. (ML services such as ChatGPT don’t reveal $E$ and $M$ and frequently change them, often silently.) Model consistency is compatible with model privacy, lack of disclosure of proprietary models and/or environments. 

In some settings, however, the consumer of a \prop’s output  $Y$ will want an exact specification of $E$ and $M$ that is replicable. PrivaLoan might, for instance, want to be able to audit a \prop used in its applications or test the \prop’s properties, such as susceptibility to adversarial inputs or hallucinations.

We define a pinned model broadly as one that includes a specification $S = (E, M)$ along with a functionality that proves that  $Y$ is the result of applying $S$ to some input $X$. The specification of $S$ can be exact, but it could also be inexact---e.g., it could be the URL for an ML service. To support the full \prop proof in privacy-preserving use cases like that in~\Cref{ex:inference}, the proof associated with a pinned model should not disclose $X$; it needs to be composable with a proof of authenticity for $X$. 

\subsection{Approaches for realizing pinned models}

Executing a model (and environment) in a TEE is one practical way to realize pinned models. Recently rolled-out support for TEEs in NVIDIA GPUs~\cite{nertney_confidential_computing} makes this approach especially viable. Another, complementary approach is to use a decentralized oracle network (DON)~\cite{breidenbach2021chainlink}. A committee of nodes in a DON could, for instance, each independently execute a model specification $S$ on an input $X$ and then reach consensus on the output  $Y$. (Or they could each execute a different model specification, a form of ensemble learning~\cite{wikipedia_ensemble_learning}.) Approaches such as zkML are also possible, but practical today only for small models~\cite{chen2024zkml}. 
 
\section{Conclusion}
\label{sec:conclusion}

\Props represent a new approach for secure, privacy-preserving access to deep-web data sources in machine learning. By enabling authenticated, privacy-preserving data pipelines, they address critical bottlenecks in data availability and model reliability within ML. \Props can ensure robust data privacy and integrity across an entire ML pipeline, from sourcing and processing to training and model execution. \Props are also flexible: They can verify data authenticity and model consistency, for instance, even in standard ML applications that don’t require privacy. In short, by combining privacy-preserving oracle systems and pinned models, \props establish a scalable pathway toward secure and reliable ML systems, unlocking the potential of deep-web data for ML capabilities.
\bigskip

{\small \colorbox{lightgray}{\begin{minipage}{0.925\textwidth}

{\textbf{ In short, \props create three significant opportunities for ML applications: \vphantom{p\^{E}}}}

\begin{enumerate}
    \item \textbf{Surfacing deep-web data}: \Props enable deep-web data—which is vastly larger than surface-web data—to be used for ML training and inference with strong privacy and integrity assurance.
    \item \textbf{Securing inference on sensitive data}: When used for inference, \props create new models of data use. Service providers can obtain ML inferences over trustworthy models with private  data. They can do so while avoiding the exposure to breaches and liability associated with direct access to sensitive data.
    \item \textbf{Constraining adversarial inputs}: For applications that rely on deep-web data inputs, authenticating inputs limits opportunities for adversarial manipulation. As a result \props can limit the impact of adversarial examples and other forms of adversarial input.
\end{enumerate}

\end{minipage}}}


\section*{Note on \Props and Blockchain Technologies}
It’s not a coincidence that most of the technical tools needed for \props saw some of their earliest production use in blockchain systems. High-assurance data delivery and application execution are especially prized in smart-contract-based blockchains, where adversaries can exploit even small vulnerabilities for quick monetary gain. Also, blockchain systems by design create a tension between transparency and privacy. Blockchains are transparent, but financial transactions usually involve sensitive data. Systems that ensure both data authenticity and privacy—such as privacy-preserving oracle systems—have sprung up to resolve this tension. 
	
\Props are not just realizable using blockchain technologies but can also be useful \textit{for} blockchain technologies. This is particularly true of \props for inference, as in~\Cref{ex:inference}. The privacy-preserving, authenticated nature of outputs makes them suitable for consumption by smart contracts. 

\section*{Acknowledgments}

Thanks to James Austgen, Lorenz Breidenbach, and Mahimna Kelkar for their helpful comments on this work. Thanks especially to Andr\'{e}s F\'{a}brega for pointing out uses for constraining adversarial inputs beyond adversarial examples.

\def\bibfont{\footnotesize}

\bibliographystyle{acm}
\bibliography{main}

\end{document}